\documentclass[twocolumn,aps,pra,showpacs,longbibliography,superscriptaddress]{revtex4-1}
\usepackage{amssymb,latexsym}
\usepackage{mathrsfs}
\usepackage{lipsum}
\usepackage{graphicx}
\usepackage[caption=false]{subfig}
\usepackage{mathtools}
\usepackage{epstopdf}
\usepackage{xcolor}
\usepackage[colorlinks, linkcolor=red, anchorcolor=green, citecolor=green, urlcolor=blue]{hyperref}
\DeclareGraphicsExtensions{.pdf,.jpg,.png,.eps}
\usepackage[toc,page,title,titletoc,header]{appendix}
\usepackage{etoolbox}
\usepackage{breqn}

\makeatletter
\let\cat@comma@active\@empty
\makeatother

\begin{document}
	
\title{Superresolution imaging of two incoherent optical sources with unequal brightnesses}

\author{Jian-Dong Zhang}
\email[]{zhangjiandong@jsut.edu.cn}
\affiliation{School of Mathematics and Physics, Jiangsu University of Technology, Changzhou 213001, China}
\author{Yiwen Fu}
\affiliation{School of Mathematics and Physics, Jiangsu University of Technology, Changzhou 213001, China}
\author{Lili Hou}
\affiliation{School of Mathematics and Physics, Jiangsu University of Technology, Changzhou 213001, China}
\author{Shuai Wang}
\affiliation{School of Mathematics and Physics, Jiangsu University of Technology, Changzhou 213001, China}
	
\date{\today}
	
\begin{abstract}
Resolving the separation between two incoherent optical sources with high precision is of great significance for fluorescence imaging and astronomical observations.
In this paper, we focus on a more general scenario where two sources have unequal brightnesses.
We give the ultimate precision 
limit with respect to separation by using the quantum Fisher information.
Through the calculation of the classical Fisher information, we analyze and compare several specific measurement schemes including direct measurement, Gaussian mode measurement and zero-photon measurement.
The results indicate that Gaussian mode measurement is the nearly optimal for a small separation.
Our work provides a positive complement to the aspect of superresolution imaging of incoherent sources.
\end{abstract}

\maketitle

\section{Introduction}

For a realistic imaging system, diffraction effects are unavoidable due to various spatial restrictions. 
In this circumstance, an ideal optical point passing through the imaging system evolves into an optical spot with a certain spatial size.
When there are two closely incoherent optical sources on the object plane, their images on the image plane will overlap extensively. 
At this time, the estimation with respect to the separation between two images is crucial, for the results can provide the position information regarding optical sources.
The estimation precision is directly proportional to the resolution of an imaging system.
A high-resolution imaging system is beneficial for many applications ranging from microscopic fluorescence imaging to macroscopic astronomical imaging.

In 2016, Tsang \emph{et al.} analyzed the ultimate precision limit with respect to the separation between two incoherent sources through the use of the quantum Fisher information \cite{PhysRevX.6.031033}.
An interesting result is that the precision limit remains constant irrespective of the separation.
In stark contrast, the precision given by classical direct measurement becomes worse as the separation tends to zero.
This indicates that, for a small separation, there exist measurement schemes which are superior to direct measurement in precision.
Since then, incoherent imaging has received a lot of attention and many super-resolved measurement schemes have been discussed.
Tsang proposed a spatial-mode demultiplexing (SPADE) measurement scheme
\cite{PhysRevA.97.023830,Tsang_2017}.
Nair \emph{et al.} showed a measurement scheme based on super localization by image inversion interferometry (SLIVER) \cite{Nair:16}.
Tham \emph{et al.} analyzed the measurement scheme using super-resolved position localisation by
inversion of coherence along an edge (SPLICE) \cite{PhysRevLett.118.070801,Bonsma-Fisher_2019}.
The corresponding proof-of-principle experiments have been also presented \cite{Tang:16,Paur:16,Larson:19,Zhou:19,Boucher:20,Tan:23}.

The above studies are a beneficial exploration for improving the resolution of imaging systems. 
However, these studies focused on two equal-brightness sources.
Related to this, some works paid attention to a more general scenario where two sources have unequal brightnesses
\cite{PhysRevA.96.062107,PhysRevA.98.012103,Prasad_2020}. 
The precision limit was analyzed through the quantum Fisher information, and the optimal decomposition modes were calculated.
More recently, Xin \emph{et al.} showed a specific super-resolved measurement scheme by utilizing fractional Hilbert transform
\cite{PhysRevA.103.052604}.
In this paper, we provide a positive complement to the aspect of separation estimation of two unequal-brightness sources.
We report the precision of Gaussian mode measurement and that of zero-photon measurement, which are two specific super-resolved measurement schemes.

The remainder of this paper is organized as follows.
Section \ref{II} introduces the fundamental model and the corresponding quantum Fisher information.
In Sec. \ref{III}, we analyze direct measurement, Gaussian mode measurement and zero-photon measurement with the method of the classical Fisher information.
The comparison between three measurement schemes is also shown.
Finally, we summarize our work in Sec. \ref{IV}.

\section{Model and corresponding precision limit of incoherent sources}
\label{II}

\begin{figure}[htbp]
	\centering
	\includegraphics[width=0.48\textwidth]{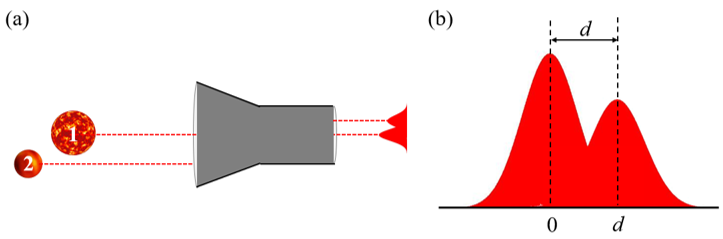}
	\caption{(a) Schematic diagram of superresolution imaging of two incoherent optical sources with unequal brightnesses. (b) The images on the image plane. The centers of two images are located at 0 and $d$.}
	\label{system}
\end{figure}

In Fig. \ref{system}, we give the model of two incoherent sources with unequal brightnesses.
Two sources passing through an optical system are imaged on the image plane.
The precision limit of a symmetrically distributed model has been discussed in Refs. \cite{PhysRevA.96.062107} and \cite{PhysRevA.98.012103}.
Throughout this paper we consider an asymmetrically distributed model in Refs. \cite{zanforlin2022optical} and \cite{PhysRevLett.127.130502}. 
Specifically, the images of two sources are located at positions of 0 and $d$, respectively.
The separation $d$ is the estimated parameter.

We first direct our attention to the precision limit of our model.
Without loss of generality, we assume that the point spread function is Gaussian and the number of photons on the image plane satisfies $\epsilon  \ll 1$.
At this time, the quantum state on the image plane can be well approximated as 
\begin{align}
\rho  \approx \left( {1 - \epsilon } \right){\rho _0} + \epsilon {\rho _1}
\end{align}
where ${\rho _0} = \left| {{\rm{vac}}} \right\rangle \left\langle {{\rm{vac}}} \right|$ is the zero-photon state, and the one-photon state can be written as 
\begin{align}
{\rho _1} = \left( {1 - q} \right)\left| 0 \right\rangle \left\langle 0 \right| + q\left| d \right\rangle \left\langle d \right|,
\end{align}
where $q$ is a parameter that describes the relative brightness of two sources.

In the coordinate representation, the amplitude profiles of two sources are found to be
\begin{align}
\left\langle {x}
{\left | {\vphantom {x {{0}}}}
	\right. \kern-\nulldelimiterspace}
{{{0}}} \right\rangle = {\left( {\frac{1}{{2\pi }}} \right)^{{1 \mathord{\left/
				{\vphantom {1 4}} \right.
				\kern-\nulldelimiterspace} 4}}}\exp \left( { - \frac{{{x^2}}}{{4}}} \right),
\label{}
\end{align}
\begin{align}
\left\langle {x}
{\left | {\vphantom {x {{d}}}}
	\right. \kern-\nulldelimiterspace}
{{{d}}} \right\rangle  = {\left( {\frac{1}{{2\pi }}} \right)^{{1 \mathord{\left/
				{\vphantom {1 4}} \right.
				\kern-\nulldelimiterspace} 4}}}\exp\left[ { - \frac{{{{\left( {x - d} \right)}^2}}}{4}} \right],
\label{}
\end{align}
where we used ${\sigma =1}$.
It is not difficult to find that the states of two sources on the image plane are not orthogonal, i.e.,
\begin{align}
\left\langle {0}
{\left | {\vphantom {0 d}}
	\right. \kern-\nulldelimiterspace}
{d} \right\rangle = \exp \left( { - \frac{{{d^2}}}{{8}}} \right) \equiv \delta \ne 0.
\label{}
\end{align}

In order to calculate the quantum Fisher information, we need to find a set of orthogonal eigenstates of $\rho_1$.
For our model, we note that $\rho_1$ only contains two states $\left| 0 \right\rangle$ and $\left| d \right\rangle$.
This feature enables us to expand $\rho_1$ in terms of two eigenstates $\left| {{\lambda _1}} \right\rangle$ and $\left| {{\lambda _2}} \right\rangle$ with nonzero eigenvalues $\lambda _1$ and $\lambda _2$.
Namely, the one-photon state can be expressed in the form of spectral decomposition
\begin{align}
{\rho _1} = {\lambda _1}\left| {{\lambda _1}} \right\rangle \left\langle {{\lambda _1}} \right| + {\lambda _2}\left| {{\lambda _2}} \right\rangle \left\langle {{\lambda _2}} \right|,
\end{align}
where $\left\langle {\lambda _1}
{\left | {\vphantom {{\lambda _1} {\lambda _2}}}
	\right. \kern-\nulldelimiterspace}
{\lambda _2} \right\rangle = 0$.
The specific forms of  eigenvalues and eigenstates can be taken as
\begin{align}
\lambda _1 &= {\frac{1}{2}}\left( 1 + \Delta \right)\\
\lambda _2 &= {\frac{1}{2}}\left( 1 - \Delta \right)\\
\left| {{\lambda _1}} \right\rangle  &= {A_1}\left| 0 \right\rangle  + {B_1}\left| d \right\rangle, \\
\left| {{\lambda _2}} \right\rangle  &= {A_2}\left| 0 \right\rangle  + {B_2}\left| d \right\rangle, 
\label{}
\end{align}
where	
\begin{align}
\Delta  = \sqrt {{{\left( {1 - 2q} \right)}^2} + {\delta ^2}\left[ {1 - {{\left( {1 - 2q} \right)}^2}} \right]} 
\label{}
\end{align}
and the parameters are given by
\begin{align}
{A_1} &= \sqrt {\frac{{\left( {1 - q} \right)\left[ {\Delta  + \left( {1 - 2q} \right)} \right]}}{{\Delta \left( {1 + \Delta } \right)}}} \\
{B_1} &= \sqrt {\frac{{q\left[ {\Delta  - \left( {1 - 2q} \right)} \right]}}{{\Delta \left( {1 + \Delta } \right)}}} \\
{A_2} &= \sqrt {\frac{{\left( {1 - q} \right)\left[ {\Delta  - \left( {1 - 2q} \right)} \right]}}{{\Delta \left( {1 - \Delta } \right)}}} \\
{B_2} &=  - \sqrt {\frac{{q\left[ {\Delta  + \left( {1 - 2q} \right)} \right]}}{{\Delta \left( {1 - \Delta } \right)}}}. 
\label{}
\end{align}

Based on the above results, the quantum Fisher information can be calculated in terms of \cite{Liu_2020}

\begin{align}
\nonumber {\cal Q} =& \sum\limits_{i = 1}^2 {\frac{1}{{{\lambda _i}}}{{\left( {\frac{{\partial {\lambda _i}}}{{\partial d}}} \right)}^2}}  + \sum\limits_{i = 1}^2 {4{\lambda _i}\left( {\frac{{\partial \left\langle {{\lambda _i}} \right|}}{{\partial d}}} \right)\left( {\frac{{\partial \left| {{\lambda _i}} \right\rangle }}{{\partial d}}} \right)} \\ 
&- \sum\limits_{i,j = 1}^2 {\frac{{8{\lambda _i}{\lambda _j}}}{{{\lambda _i} + {\lambda _j}}}} \left( {\frac{{\partial \left\langle {{\lambda _i}} \right|}}{{\partial d}}\left| {{\lambda _j}} \right\rangle } \right)\left( {\left\langle {{\lambda _j}} \right|\frac{{\partial \left| {{\lambda _i}} \right\rangle }}{{\partial d}}} \right).
\label{}
\end{align}
The ultimate precision limit is the reciprocal of  square root of the quantum Fisher information.

\begin{figure}[htbp]
	\centering
\includegraphics[width=0.47\textwidth]{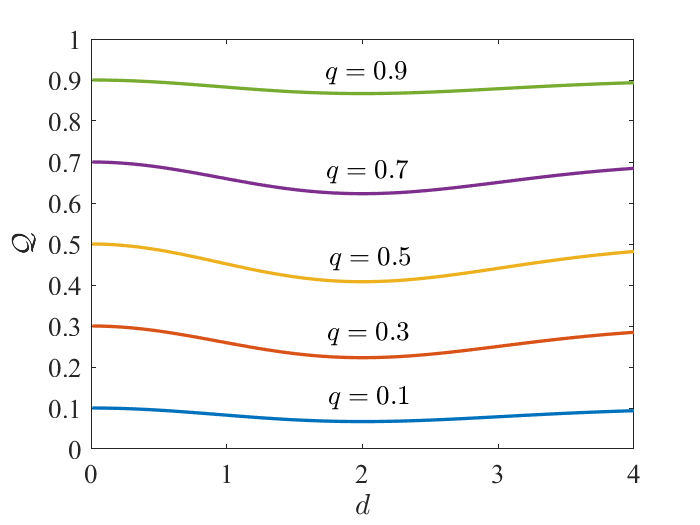}
	\caption{The quantum Fisher information as a function of the separation with different values of $q$.}
	\label{QFI}
\end{figure}

In order to intuitively observe the quantum Fisher information against the separation, we provide the result in Fig. \ref{QFI}.
It turns out that the quantum Fisher information does not remain constant when the separation changes. 
The minimum value occurs at approximately $d=2$ for any value of $q$.
However, the fluctuation degree of the quantum Fisher information is different for different values of $q$.
In addition, the quantum Fisher information increases with the increase of $q$. 
This is due to the fact that the second source becomes brighter as the value of $q$ increases, and this source carries information about the separation.

\begin{figure*}[htbp]
	\centering
\includegraphics[width=0.45\textwidth]{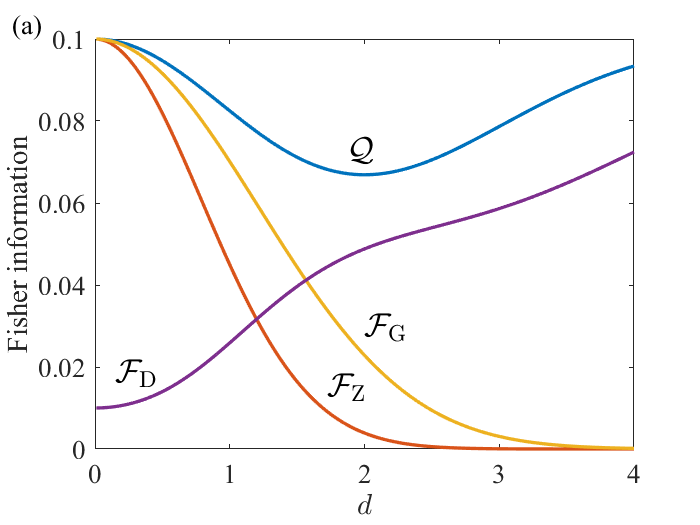}
\includegraphics[width=0.45\textwidth]{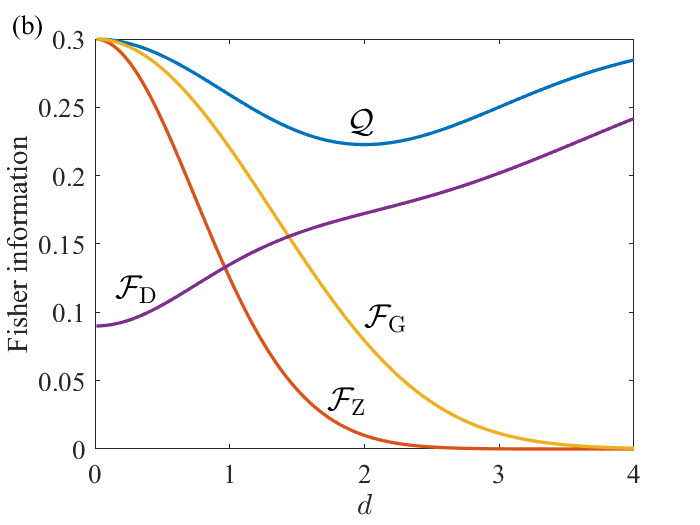}
\includegraphics[width=0.45\textwidth]{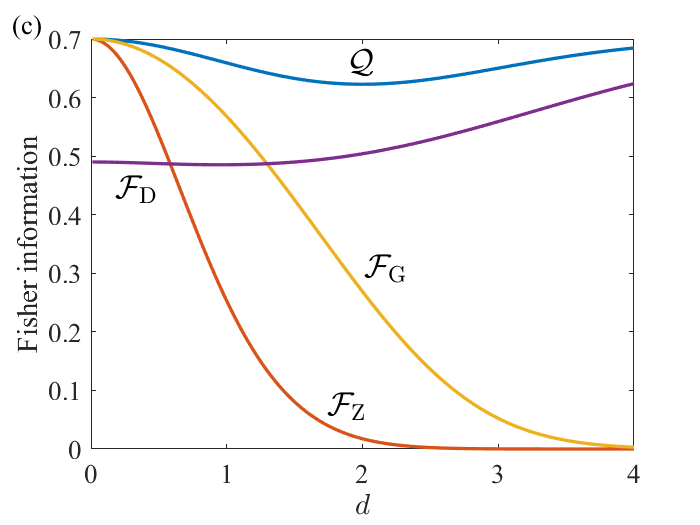}
\includegraphics[width=0.45\textwidth]{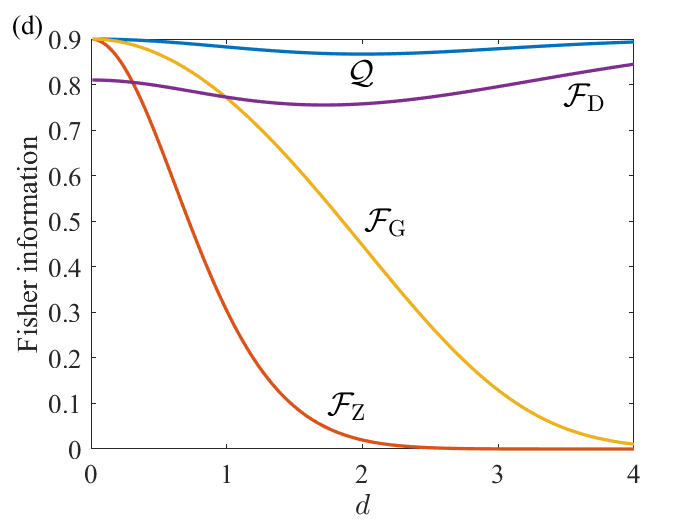}
	\caption{The classical Fisher information of three measurement schemes as a function of the separation, (a) $q =0.1$, (b) $q =0.3$, (c) $q =0.7$, (d) $q =0.9$.}
	\label{CFI}
\end{figure*}

\section{Calculation and comparison of specific measurement schemes in precision}
\label{III}

In this section, we turn our attention to analyze two specific measurement schemes including Gaussian mode measurement and zero-photon measurement.
As a benchmark, we first analyze the classical direct measurement scheme.
On the image plane, the intensity profile reads
\begin{align}
\nonumber \Lambda  &= \left\langle x \right|{\rho _1}\left| x \right\rangle  \\
&= \frac{{1 - q}}{{\sqrt {2\pi } }}\exp \left( { - \frac{{{x^2}}}{2}} \right) + \frac{q}{{\sqrt {2\pi } }}\exp \left[ { - \frac{{{{\left( {x - d} \right)}^2}}}{2}} \right].
\label{}
\end{align}
Since the intensity profile is continuous in the coordinate representation, the corresponding classical Fisher information can be calculated according to
\begin{align}
{{\cal F}_{\rm D}} = \int_{ - \infty }^\infty  {dx\frac{1}{\Lambda }{{\left( {\frac{{\partial \Lambda }}{{\partial d}}} \right)}^2}}. 
\label{}
\end{align}

Gaussian mode measurement is a binary scheme belonging to the SPADE measurement scheme.
Specifically, a single-mode optical fiber is aligned with the position of 0 and resolves the spatial mode of photons on the image plane. 
Correspondingly, only photons in Gaussian mode can be coupled into the fiber.
The probabilities of identifying Gaussian and non-Gaussian modes are denoted as ${P_{\rm G}}$ and $1 - {P_{\rm G}}$, respectively.
The probability ${P_{\rm G}}$ is given by  
\begin{align}
{P_{\rm G}} = \left\langle 0 \right|{\rho _1}\left| 0 \right\rangle  = 1 + q\left[ {\exp \left( { - \frac{{{d^2}}}{4}} \right) - 1} \right],
\label{}
\end{align}
and the corresponding classical Fisher information turns out to be 
\begin{align}
{{\cal F}_{\rm G}} = \frac{1}{{{P_{\rm G}}}}{\left( {\frac{{\partial {P_{\rm G}}}}{{\partial d}}} \right)^2} + \frac{1}{{1 - {P_{\rm{G}}}}}{\left[ {\frac{{\partial \left( {1 - {P_{\rm{G}}}} \right)}}{{\partial d}}} \right]^2}.
\label{}
\end{align}

Zero-photon measurement is also a binary scheme, which belongs to the SLIVER measurement scheme.
In this scheme, we need to spatially separate the photons on the image plane according to symmetric and antisymmetric parts.
This can be achieved through the use of a Mach-Zehnder interferometer embedded with a Dove prism \cite{Nair:16}.
Then we record the photons originating from the antisymmetric part.
The probabilities of identifying zero and non-zero photons are denoted as ${P_{\rm Z}}$ and $1 - {P_{\rm Z}}$, respectively.
This measurement can be achieved by using a Gm-APD that can only distinguish the presence or absence of photons.
By using semiclassical method (see Appendix for details), the probability ${P_{\rm Z}}$ is found to be
\begin{align}
{P_{\rm{Z}}} = \frac{2}{{2 + q\left[ {1 - \exp \left( { - {{{d^2}} \mathord{\left/
						{\vphantom {{{d^2}} 2}} \right.
						\kern-\nulldelimiterspace} 2}} \right)} \right]}}.
\label{e22}
\end{align}
Accordingly, the classical Fisher information is given by
\begin{align}
{{\cal F}_{\rm Z}} = \frac{1}{{{P_{{\rm{Z}}}}}}{\left( {\frac{{\partial {P_{{\rm{Z}}}}}}{{\partial d}}} \right)^2} + \frac{1}{{1 - {P_{\rm{Z}}}}}{\left[ {\frac{{\partial \left( {1 - {P_{\rm{Z}}}} \right)}}{{\partial d}}} \right]^2}.
\label{}
\end{align}
The specific precision
of each measurement scheme is the reciprocal of square root of the corresponding classical Fisher information.

As a comparison among the above measurement schemes, we give the dependence of the classical Fisher information on the separation, as shown in Fig. \ref{CFI}.
The results indicate that, for a small separation, there is a significant difference between the classical Fisher information of the direct measurement and the quantum Fisher information.
On the other hand, the classical Fisher information of Gaussian mode measurement and that of zero-photon measurement can saturate with the quantum Fisher information when the separation approaches 0.
By comparison, the classical Fisher information of Gaussian mode measurement is superior to that of zero-photon measurement.
This suggests that Gaussian mode measurement is the approximately optimal scheme for a small separation.

\begin{figure}[htbp]
	\centering
\includegraphics[width=0.48\textwidth]{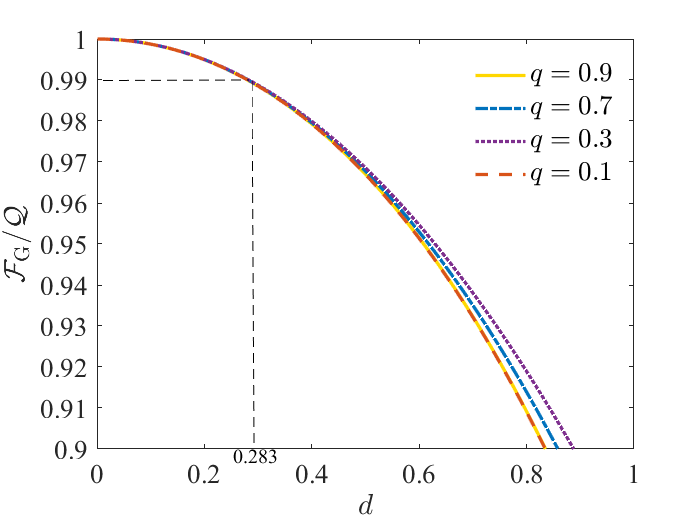}
	\caption{The ratio of classical Fisher information of Gaussian mode measurement to the quantum Fisher information as a function of the separation with different values of $q$.}
	\label{optimal}
\end{figure}

To quantify the optimality, in Fig. \ref{optimal} we show the ratio of classical Fisher information of Gaussian mode measurement to the quantum Fisher information.
This ratio does not change significantly with the value of $q$, especially within the interval of $0\le d \le 0.5$.
In particular, when the separation satisfies $d<0.283$, the classical Fisher information of Gaussian mode measurement can exceed 99\% of the quantum Fisher information.
That is, Gaussian mode measurement can maintain approximately optimal precision over a considerable range of the separation.

It should be noted that the use of Gaussian mode measurement in this paper is for the sake of simplicity in experimental implementation.
In theory, one can further improve the classical Fisher information if the measured modes are increased.
A direct result is that the measurement can provide approximately optimal precision over a larger separation range.

\section{Conclusion}
\label{IV}
In summary, the separation estimation problem in the scenario of two unequal-brightness incoherent sources was addressed.
We calculated the quantum Fisher information of our model.
Three measurement schemes, direct measurement, Gaussian mode measurement and zero-photon measurement, were analyzed through the use of the classical Fisher information.
Gaussian mode measurement was found to be the approximately optimal scheme, and the corresponding classical Fisher information can
exceed 99\% of the quantum Fisher information as long as $d \le 0.283$.
These results may be beneficial for many applications such as fluorescence imaging and astronomical observations.

\section*{Acknowledgment} 
This work was supported by the Program of Zhongwu Young Innovative Talents of Jiangsu University of Technology (20230013).

\appendix

\section*{Appendix} 
In this section, we provide the calculation of the probability $P_{\rm Z}$.
According to the semiclassical theory, the complex amplitude of the states on the image plane can be written as
\begin{align}
E\left( x \right) = {C_1}\psi \left( x \right) + {C_2}\psi \left( {x - d} \right).
\label{}
\end{align}	
The amplitudes are circular-complex Gaussian random variables satisfying
\begin{align}
&{\mathbf E}\left[ {{C_1}} \right] = {\mathbf E}\left[ {{C_2}} \right] = 0,\\
&{\mathbf E}\left[ {{C_1}{C_2}} \right] = {\mathbf E}\left[ {C_1^ * {C_2}} \right] = {\mathbf E}\left[ {{C_1}C_2^ * } \right] = 0,\\
&{\mathbf E}\left[ {C_1^ * {C_1}} \right] = 1 - q,\\
&{\mathbf E}\left[ {C_2^ * {C_2}} \right] = q.
\label{}
\end{align}

For SLIVER measurement, the complex amplitude of antisymmetric part can be calculated from
\begin{align}
{E_{\rm a}}\left( x \right) = \frac{{E\left( x \right) - E\left( { - x} \right)}}{2} = \frac{{{C_2}}}{2}\left[ {\psi \left( {x - d} \right) - \psi \left( {x + d} \right)} \right],
\label{}
\end{align}	
where we used the
symmetry of the point spread function, i.e.,
\begin{align}
{\psi \left( x \right)}  = {\psi \left( -x \right)}.
\label{}
\end{align}

Further, the mean photon number in the antisymmetric part is given by
\begin{align}
{N_{\rm a}} = \int_{ - \infty }^\infty  {{{\left| {{E_a}\left( x \right)} \right|}^2}dx}  = \frac{{q}}{2}\left[ {1 - \exp \left( { - \frac{{{d^2}}}{2}} \right)} \right].
\label{e31}
\end{align}	
Since the photon number satisfies Bose-Einstein distribution, the probability of zero-photon events is found to be
\begin{align}
{P_0} = \frac{1}{{1 + {N_{\rm a}}}}.
\label{e32}
\end{align}	
Finally, the probability $P_{\rm Z}$ in Eq. (\ref{e22}) is obtained by combining Eqs. (\ref{e31}) and (\ref{e32}).

\end{document}